\documentclass[aip,jmp,reprint,apl]{revtex4-1}
\usepackage{graphicx}
\usepackage{dcolumn}
\usepackage{bm}
\usepackage[dvips]{color}
\usepackage{verbatim}

\begin{document}


\title{Stable gain-switched thulium fiber laser with 140 nm tuning range}

\author{Fengqiu Wang}
\email{fwang@nju.edu.cn}
\author{Yafei Meng}
\affiliation{School of Electronic Science and Engineering and Collaborative Innovation Center of Advanced Microstructures, Nanjing University, Nanjing
210093, China}
\author{Edmund Kelleher}
\email{edmund.kelleher08@imperial.ac.uk}
\affiliation{Femtosecond Optics Group, Department of Physics, Imperial College London, SW7 2AZ, UK}
\author{Guoxiang Guo}
\author{Yao Li}
\author{Yongbing Xu}
\affiliation{School of Electronic Science and Engineering and Collaborative Innovation Center of Advanced Microstructures, Nanjing University, Nanjing
210093, China}
\author{Shining Zhu}
\affiliation{National Laboratory of Solid State Microstructures and School of Physics, Nanjing University, Nanjing 210093, China}

\begin{abstract}
We demonstrate a gain-switched thulium fiber laser that can be continuously tuned over 140 nm, while maintaining stable nanosecond single-pulse operation. To the best of our knowledge, this system represents the broadest tuning range for a gain-switched fiber laser. The system simplicity and wideband wavelength tunability combined with the ability to control the temporal characteristics of the gain-switched pulses mean this is a versatile source highly suited to a wide range of applications in the eye-safe region of the infrared, including spectroscopy, sensing and material processing, as well as being a practical seed source for pumping nonlinear processes.

\end{abstract}

\maketitle

Sustained interest in the development of thulium-doped fiber based technologies is motivated by the numerous applications that benefit from laser sources in the two micron spectral region.\cite{Geng-2014,Jackson-NP-2012} The strong absorption at $\sim$1.93 $\mu$m -- corresponding to the first overtone vibration of the OH molecule \cite{Dausinger-book-2004} -- can be exploited in the medical sector for laser surgery,\cite{Matsuoka-LSM-1999,Sumiyoshi-1999-review} while an atmospheric transmission window in this region means high-brightness sources with coincident emission bands are attractive for range finding, remote sensing and free-space optical communications.\cite{Ebrahim-Zadeh-book-2008} To meet industrial demands, research efforts continue to focus on thulium-doped fiber for the development of both pulsed and continuous-wave two micron laser technologies due to its extremely broad emission band (approx. 1.8 $\mu$m -- 2.1 $\mu$m) that supports wideband tunable systems,\cite{Nelson-1995,Kieu-2009,McComb-2010} in addition to the intrinsic benefits of a fiber architecture including: excellent beam quality; thermal robustness; compact alignment free operation; and power scalability.\cite{Wang-NatureNanotechnol-2008,Stutzki-OL-2014,Stutzki-OL-2015}

Gain-switching of a laser -- by fast modulation of the pump power\cite{Ho-APL-1978,Zayhowski-1989} -- is perhaps the simplest and most robust way to generate a train of short pulses and has been applied to fiber lasers operating across the near-infrared (1--2~$\mu$m).\cite{Larsen-OE-2014,Tsai-AIP-2011,Xu-invited-2013,Jiang-OL-2007,Jackson-QE-1998} By appropriate choice of the pump pulse-energy and duration, mutliple relaxation oscillations can be prevented, mitigating temporal instabilities and leading to the generation of stable Gaussian-like nanosecond pulses.\cite{Xu-invited-2013,Jiang-OL-2007} As both the repetition rate and pulse duration can be controlled by the pump characteristics, gain-switched fiber lasers exhibit greater temporal flexibility compared to other pulse generation techniques, including Q-switching and mode-locking, where the dynamics of both the gain and an intra-cavity loss modulator, as well as dispersion of the laser cavity, play dominant contributions to pulse-shaping.\cite{Agrawal-book-2004} Another desirable feature of gain-switched lasers is that they can be readily synchronized with an external reference source. Despite significant technical progress in recent years, including the demonstration of a pulse duration as short as $\sim$1.5~ns and peak powers in excess of 100~kW,\cite{Ding-SPIE-2011,Tang-LPL-2013} gain-switched thulium fiber lasers have typically adopted a design based on the use of fiber Bragg gratings (FBGs) forming the cavity end mirrors,\cite{Hou-QE-2014,Swiderski-QE-2014,Xu-JOSAB-2014,Swiderski-OL-2013,Hou-JOSAB-2013,Simakov-OE-2011} and thus offer only narrow-band operation. Although the reflection spectrum of a FBG can be tuned by applying a strain or a temperature change to the component,\cite{Jeon-OE-2010} the degree of tunability achievable does not permit full exploitation of the ultra-wideband emission of thulium-doped fiber. Consequently, gain-switched systems reported to date have supported a limited range of wavelength tuning; yet, the ability to tune the operation wavelength of a laser is of vital importance for many applications, including spectroscopy and sensing.\cite{Coldren-JLT-2004}

In this letter, we demonstrate a broadly-tunable gain-switched thulium fiber laser based on a wideband reflective diffraction grating mirror. By utilizing such a simple and FBG-free cavity design we achieve an extremely wide range of wavelength tuning, approximately 140~nm (from 1860 -- 2000 nm), to the best of our knowledge the broadest tuning of a gain-switched fiber laser, while generating stable single nanosecond-scale pulses with controllable duration at a selectable repetition rate. Due to the reliability of pulsed operation, as demonstrated by the high side-mode suppression ratio (SMSR) of the electrical spectrum, we anticipate that this source will prove a popular cost-effective solution for industrial applications of pulsed lasers in the two micron spectral region.

Figure 1 illustrates the in-band-pumped gain-switched laser setup. A 1550 nm laser diode is directly modulated by a pulsed current driver (Newport LDP-3830) and subsequently amplified by a high power erbium-doped fiber amplifier (EDFA). The pump pulse is coupled into the cavity via a wavelength-division-multiplexer (WDM). The gain-switched laser is formed by a linear cavity and 2.4 m of thulium-doped fiber, with 9/125 core/cladding geometry (Coractive Inc.), comprising the gain medium. The diffraction grating end mirror has a blaze wavelength of 1600 nm, and a groove density of 600/mm. A fiber coupled collimator collimates the light before irradiating the grating. A Littrow grating configuration is used, where a portion of the diffracted beam is reflected back into the laser cavity.\cite{Kneis-OL-2015} The other end of the cavity is formed by a fiber patchord with a partially reflective coating (20\% transmission and 80\% reflection) that has a broad bandwidth extending from 1800 -- 2100 nm. The total cavity length is approximately 7 m, including a free-space section between the fiber collimator and the grating. The laser output is connected to an optical spectrum analyzer (Yokogawa AQ6375), a fast oscilloscope (Agilent DSO-X3052A), and a radio frequency spectrum analyzer (R\&S FSV 30) for characterization and analysis of the laser performance.

\begin{figure}
\centering\includegraphics[width=1\columnwidth]{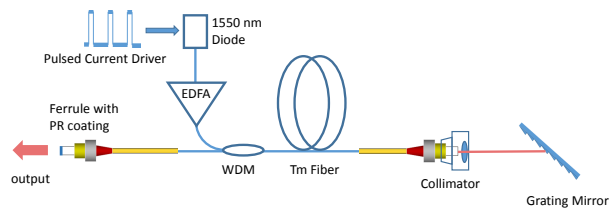}
\caption{Experimental setup of the broadly tunable gain-switched thulium fiber laser utilizing a diffraction grating based end mirror.}
\label{fig1}
\end{figure}

We first investigate the laser's threshold and pump conditions for single-pulse operation at a wavelength of 1940 nm. When the pump pulse is set to 200 ns duration and 25 kHz repetition rate, the threshold for cw laser operation is found to be $\sim$370 mW. When the pump power is greater than 415 mW, the laser could operate in a stable, single-pulse regime. The maximum available pump power is 575 mW, limited by the EDFA used. For investigating the broadband spectral characteristics, the pump power is held at a fixed value of 530 mW (corresponding to a pump pulse energy of 21 $\mu$J), while the grating mirror is tuned. Fig. 2 shows the spectra at different operating wavelengths. The full width half maximum (FWHM) spectral bandwidth for the gain-switched operation is $\sim$0.5 nm as revealed by higher resolution measurement. The spectra stay stable with a high optical signal-to-noise ratio (OSNR) of $>$ 40 dB and a spectral flatness of $\sim$ 3 dB. Figure 3 illustrates the output power measured after the partially reflective end mirror as a function of operating wavelength. A second WDM is used outside the cavity to remove residual 1550 nm pump light. The output power ranges from 0.8 - 3.2 mW, with higher efficiency achieved for the 1900 - 1960 nm band, consistent with the emission profile of the thulium-doped fiber used and a roll-off in the reflection of the coated output mirror below 1900 nm. Figure 3 also shows the measured pulse duration in the stable single-pulse regime, at a fixed pump power of 530 mW. It should be noted that with this pump condition, a single pulse is readily achieved across the whole wavelength tuning range. The pulse durations at different wavelengths varies in the range 350 - 850 ns, and scales inversely with the output power.
\begin{figure}
\centering\includegraphics[width=1\columnwidth]{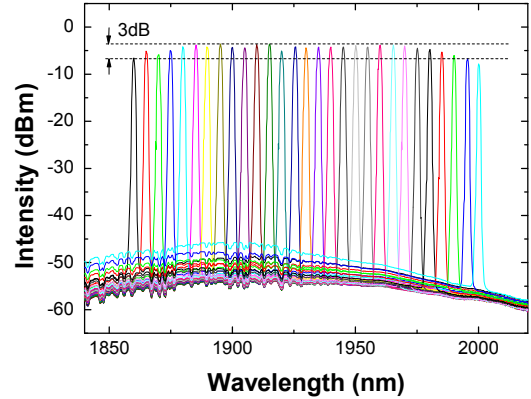}
\caption{Spectra of the gain-switched thulium fiber laser, tuning range 1860 - 2000 nm, at a fixed pump power of 530 mW.}
\label{fig2}
\end{figure}

\begin{figure}
\centering\includegraphics[width=1\columnwidth]{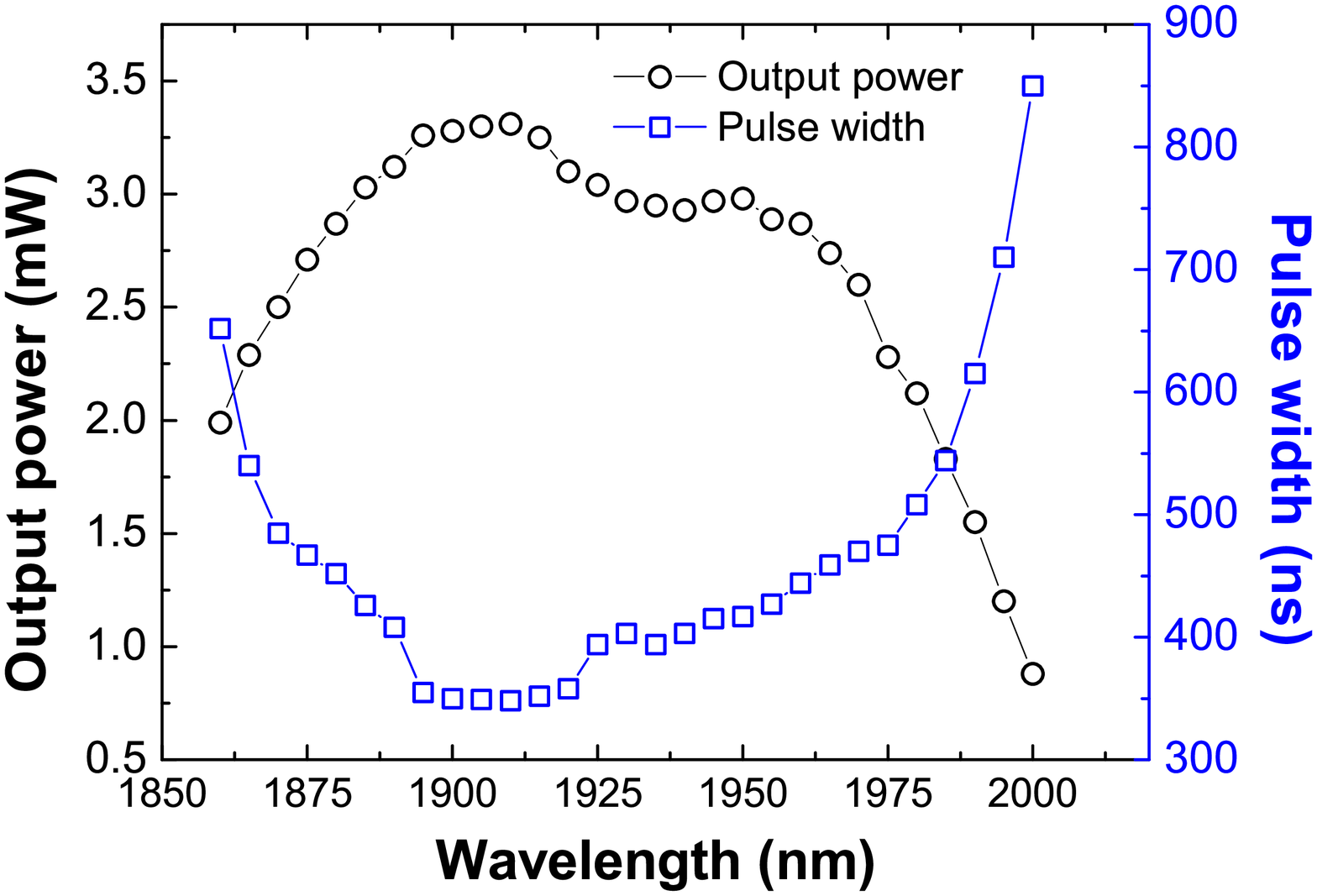}
\caption{Output power and pulse width as a function of operating wavelength across the tuning range, at a fixed pump power of 530 mW.}
\label{fig3}
\end{figure}
The temporal characteristics of the laser output under the same pump conditions is illustrated in Figure 4. The oscilloscope traces at 1875 nm, 1935 nm and 1975 nm highlight the clean single-pulse operation, distinct from a chaotic oscillatory output indicative of multiple oscillation relaxations of the gain typical of systems with slow pump modulation.\cite{Zhang-OE-2005} The pulse duration is slightly elongated with increasing operating wavelength, as shown in Fig. 3.  While there are previous reports of stable gain-switched thulium fiber lasers, the criteria for stable operation of such lasers has not been fully established. Here, stable operation of our gain-switched laser is confirmed by direct measurement of the the RF spectrum. Across the entire tuning range of the laser, the SMSR of the fundamental harmonic frequency is $\sim$40 dB (measured with a resolution bandwidth of 100 Hz), a value typically used to quantify stable Q-switched systems.\cite{Popa-APL-2011}

\begin{figure}
\centering\includegraphics[width=1\columnwidth]{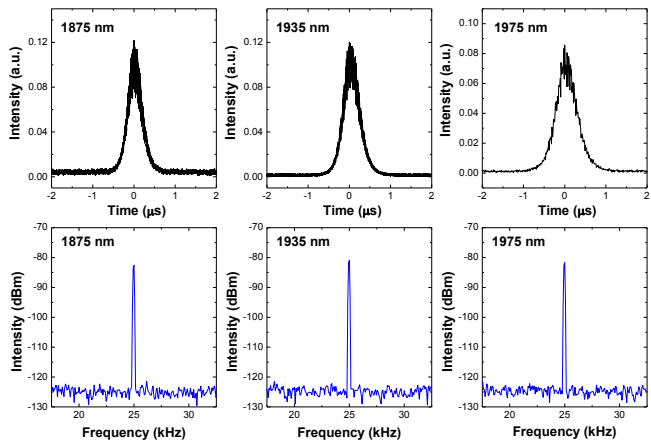}
\caption{ Pulse temporal profile and RF spectra at representative operating wavelengths within the tuning range.}
\label{fig4}
\end{figure}

It is well known that in a gain-switched laser the output pulse duration is a function of the pump pulse energy, as an increasing pump energy reduces the pulse build-up time.\cite{Xu-invited-2013} Similarly, elongation of the laser cavity, increases the pulse build-up time, and consequently the pulse duration. This trend is exemplified in Fig. 5(a) at a wavelength of 1930 nm for a pump configuration of 200 ns at 25 kHz repetition rate. It can be seen that the measured gain-switched pulse width monolithically decreases from 625 ns to 350 ns, at a rate of $\sim$50 ns/$\mu$J. In our laser, the output pulse duration is found to be appreciably longer than the pump pulses. This is attributed to the relatively long cavity length.\cite{Jiang-OL-2007} Fig. 5(b) shows the pulse duration dependence on cavity length. By optimizing the pump pulse durations, i.e. using 1 $\mu$s pump duration and 5 kHz repetition rage, at a cavity length of 5 m, the pulse duration can be reduced down to $\sim$175 ns. We note that this is compatible with pulse widths reported in other gain-switched thulium lasers with $\sim$5 m cavity and similar pump conditions.\cite{Jiang-OL-2007,Xu-JOSAB-2014} Shorter pulse durations are expected if the cavity length can be further reduced with the use of highly-doped thulium fiber. To compare the experimental measurements with the performance expected for conventional FBG-based cavities, we numerically investigate the pulse width dependence on cavity length using a simplified theoretical model of the temporal dynamics for in-band pumped gain-switched thulium fiber lasers, neglecting axial variations along the fiber.\cite{Zhou-CPB-2013} The cavity and pump parameters are set according to the experimental conditions and other parameters such as absorption and emission cross sections are taken from the literature.\cite{Agger-OE-2006,Xu-JOSAB-2014,Zhou-CPB-2013} The results are shown by a dashed line in Fig. 5(b). Good qualitative agreement between simulation and experiment is obtained, suggesting the pulse formation dynamics are not strongly influence by the inclusion of the grating mirror compared to an FBG-based design. This is expected, given that the grating mirror has a similar passband width ($\sim$0.5 nm) to an FBG device, and gain-switched lasers with long pulses ($>$10 ns) generally operate in a quasi-linear regime.\cite{Xu-JOSAB-2014} Fig. 5(c) illustrates the flexibility of the gain-switching approach as a method for generating nanosecond-scale pulses: by simply changing the pump repetition rate from 5 kHz to 25 kHz, stable single-pulses can be consistently obtained. The output can be further amplified to an average power level of $\sim$2 W, using a single-stage 793 nm pumped (total pump power 12 W) double-clad thulium fiber amplifier, leading to kilo-Watt level peak powers and pulse-energies approaching 0.5 mJ.

\begin{figure}
\centering\includegraphics[width=0.9\columnwidth]{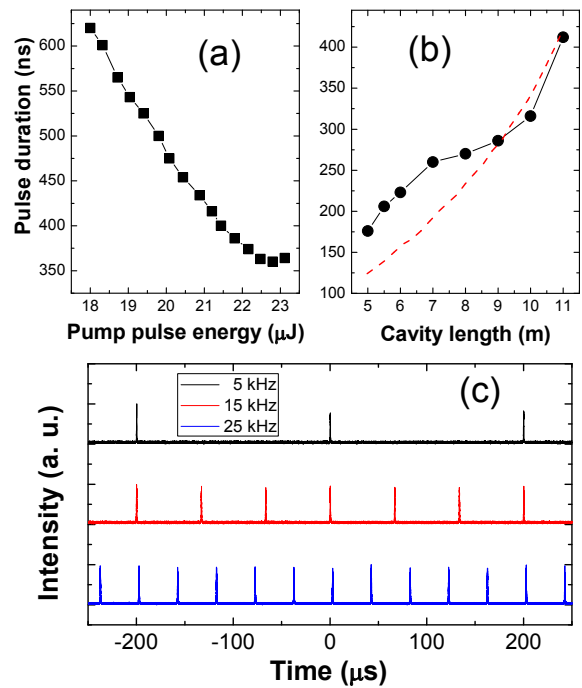}
\caption{(a) Pulse duration as a function of pump pulse energy (25 kHz, 200 ns), 7 m cavity length, 1930 nm operation. (b) Pulse duration as a function of cavity length at 1930 nm, under 5 kHz, 1 $\mu$s, and 77 $\mu$J pump pulse energy. Dash line denotes numerical results from model in Ref. \citenum{Zhou-CPB-2013} . (c) Oscilloscope traces of operation at different pump repetition rates, 1 $\mu$s, and 77 $\mu$J pump pulse energy.}
\label{fig5}
\end{figure}

In conclusion, we have proposed and demonstrated a FBG-free gain-switched thulium-fiber laser, utilizing a diffraction grating mirror. Broadly tunable gain-switched operation over an ultra-broad tuning range of 140 nm is achieved. The laser delivers stable single-pulses throughout the entire tuning range, with a controllable output pulse duration at a selectable repetition rate. This robust and simple laser geometry effectively harnesses the unique broadband capability of thulium-doped fiber, and may prove a versatile and low-cost source for industrial applications including, sensing and spectroscopy, surgery, and material processing, or a flexible seed source in master oscillator power amplifiers and as a pump scheme for nonlinear wavelength conversion to the mid-IR.

We acknowledge funding from National Key Basic Research Program of China (2014CB921101), National Natural Science Foundation of China (61378025, 61450110087, 61327812), Jiangsu Province Shuangchuang Team program, and State Key Laboratory of Advanced Optical Communication Systems Networks, China.

\end{document}